\documentclass[apj]{emulateapj}

\usepackage{amssymb, amsmath, graphicx, enumerate}
\usepackage{aas_macros, natbib, ulem}

\usepackage{times}

\shorttitle{Translating the GWB and astrophysical measurements}
\shortauthors{Simon and Burke-Spolaor}

\newcommand{\myemail}{simon3@uwm.edu}

\newcommand{\hc}{h_{\rm c}}
\newcommand{\ayr}{A_{\rm yr}}
\newcommand{\msun}{\rm M_\odot}
\newcommand{\mbulge}{M_{\rm bulge}}
\newcommand{\mmb}{\mbox{$M_{\bullet}$-$\mbulge$}}
\newcommand{\fbulge}{f_{\rm bulge}}
\newcommand{\fpair}{f_{\rm pair}}
\newcommand{\eg}{e.\,g.}
\newcommand{\strainspec}{h_{\rm c} (f)}
\newcommand{\dist}{D_{\rm c}}
\newcommand{\tstall}{T_{\rm stall}}
\newcommand{\fgearth}{f}
\newcommand{\fgrest}{f_{\rm r}}
\newcommand{\alphah}{\alpha_{\rm h}}

\begin{document}



\title{Constraints on Black Hole/Host Galaxy Co-evolution and Binary Stalling Using Pulsar Timing Arrays}


\author{Joseph Simon\altaffilmark{1} and 
Sarah Burke-Spolaor\altaffilmark{2,3}}
\affil{$^{1}$Center for Gravitation, Cosmology and Astrophysics, University of Wisconsin Milwaukee, PO Box 413, Milwaukee WI 53201, USA}
\affil{$^{2}$National Radio Astronomy Observatory, PO Box O, 1003 Lopezville Rd, Socorro, NM 87801-0387, USA}
\affil{$^{3}$Jansky Fellow}
\email{\myemail}

\begin{abstract}
Pulsar timing arrays are now setting increasingly tight limits on the gravitational wave background from binary supermassive black holes.
But as upper limits grow more constraining, what can be implied about galaxy evolution? 
We investigate which astrophysical parameters have the largest impact on strain spectrum predictions and provide a simple framework to directly translate between measured values for the parameters of galaxy evolution and PTA limits on the gravitational wave background of binary supermassive black holes. 
We find that the most influential observable is the relation between a host galaxy's central bulge and its central black hole, $\mmb$, which has the largest effect on the mean value of the characteristic strain amplitude. However, the variance of each prediction is dominated by uncertainties in the galaxy stellar mass function.
Using this framework with the best published PTA limit, we can set limits on the shape and scatter of the $\mmb$ relation. We find our limits to be in contention with strain predictions using two leading measurements of this relation. We investigate several possible reasons for this disagreement.
If we take the $\mmb$ relations to be correct within a simple power-law model for the gravitational wave background, then the inconsistency is reconcilable by allowing for an additional ``stalling'' time between a galaxy merger and evolution of a binary supermassive black hole to sub-parsec scales, with lower limits on this timescale of $\sim 1-2$\,Gyr.
\end{abstract}

\section{Introduction}
Pulsar timing arrays (PTAs) are beginning to place meaningful constraints on the gravitational wave background (GWB) in the nanoHz-$\mu$Hz gravitational wave band. In particular, recent upper limits have cut into the range of predicted strengths of the GWB from supermassive black hole (SMBH) binaries \citep{EPTA+15,nanograv+15,ppta+15}.

The characteristic strain spectrum of the GWB, $\strainspec$, is the quadrature sum of individual sources emitting in a certain GW frequency band. 
The first predictions of the GWB signal in the PTA band were made primarily using semi-analytical dark matter halo simulations \citep[\eg][]{jaffebacker,wyitheloeb}. These have been updated with full cosmological simulations (i.e. the Millennium simulation \citep{Sesana+09, Ravi+12}).
In recent years it has been recognized that the dominant source of emission in the PTA band may come from the $z\lesssim2$ binary population \citep{sesana+08} and efforts to predict $\strainspec$ have been based on various combinations of simulated and observed components of cosmological galaxy evolution. In fact, a number of the parameters that make up the astrophysical GWB are well-constrained by observation.
Thus, the focus has turned toward modeling the GWB based on local observables \citep{sesana13,mcwilliams+14,ravi+15}. Current work attempts to understand the effects that both under-efficient inspiral (binary ``stalling'') and super-efficient inspiral (i.\,e.\ binary coupling with the galaxy core environment through the PTA band) might have on $\strainspec$ \citep{sesana13CQG,ravi+14,huerta+15,sampson+15}. Inspiral efficiency affects both the predicted amplitude and the spectral shape of the of GWB, but it is unclear how the effects described above map into galaxy evolution parameters.

In this work, we aim to provide an accessible ``translation'' of GW strain directly to the parameters contributing to the largest uncertainties in the GWB spectrum. Our base assumptions are
1) circular SMBH binaries, and 2) dominantly GW-driven inspiral through the nHz--$\mu$Hz GW band (i.\,e.~a single power-law GWB). Section \ref{sec:model} lays out the mathematical preliminaries of galaxy evolution parameters that contribute to the strain spectrum. These parameters provide the basis for our construction of the astrophysical GWB.
Section \ref{sec:obscon} describes the state-of-the-art observational constraints on these parameters from the literature used in this work.
Section \ref{sec:results} presents how each astrophysical parameter effects the strain amplitude, specifically exploring a direct mapping to the $\mmb$ relation.
Finally, Sec.\,\ref{sec:ptacon} explores an example use-case for the results of this paper.

\section{Model Of PTA Strain Spectrum}\label{sec:model}
The polarization and sky-averaged strain from one circular binary system is 
\begin{equation}\label{eq:hrms}
    h_{s} = \sqrt{\frac{32}{5}} \left(\frac{G M_{c}}{c^{3}} \right)^{5/3} \frac{\left( \pi \fgrest \right)^{2/3} c}{\dist}~,
\end{equation}
\citep{thorne87}, where $M_{c} = (M_{1} M_{2})^{3/5}/(M_{1} + M_{2})^{1/5}$ is the chirp mass of the binary, $\dist$ is the proper (co-moving) distance to the binary \citep[\eg][]{petersmathews63}. $\fgrest$ is the frequency of the GWs emitted in the rest frame of the binary, where Earth-observed frequency \mbox{$\fgearth=\fgrest/(1+z)$} for a circular binary.
In the absence of external influence from the galactic environment, the orbit of such a system changes due to the emission of gravitational radiation at a rate \citep{petersmathews63}:
\begin{equation}\label{eq:dtdf}
    \frac{dt}{d\fgearth} = \frac{5}{96} \left( \frac{c^{3}}{G M_{c}} \right)^{5/3} \frac{\pi^{8/3}}{\fgearth^{11/3}}~.
\end{equation}

As in \citet{sesana+08}, we construct the characteristic strain spectrum for binary SMBHs as 
\begin{equation}\label{eq:strain}
h_{c}^{2}(\fgearth) = \int \int \int \frac{d^{4}N}{dz ~dM ~dq ~d(\textnormal{ln} \fgearth)} h_{s}^{2} ~dz ~dM ~dq~,
\end{equation}
where $d^{4}N$ is the number of binaries in a given redshift range $dz$, mass range $dM$, and mass ratio range $dq$, which are emitting in a given GW frequency range $d({\rm ln} \fgearth)$. $h_{s}^{2}$ is the polarization and sky-averaged strain from each binary in that range, Eq.\,\ref{eq:hrms}. 
Note that we expand chirp mass here into mass, $M$, and mass ratio, $q$, with $q \leq 1$: $M_c = M (q^{3}/(1+q))^{1/5}$.
The contents of the triple integral encapsulate the total number of binaries in a given frequency interval which, due to the constraints of human lifetime and measurement capabilities, are predicted based on the evolutionary properties of the galaxies that host the relevant SMBHs:
\begin{equation}
 \frac{d^{4}N}{dz ~dM ~dq ~d(\textnormal{ln}\fgearth)}  = \frac{d^{3}n}{dz ~dM ~dq} ~ \frac{dV_{c}}{dz} ~ \frac{dz}{dt}	~ \frac{dt}{d(\textnormal{ln}\fgearth)}~.
 \label{eq:d4N}
\end{equation}
In this expression, $d^{3}n$ is the number density of binaries in a given redshift, mass, and mass ratio range ($dz$, $dM$, and $dq$, respectively). The resulting conversion takes the number of binaries per co-moving volume shell, $dV_{c}$, and converts it to the number of binaries per earth-observed gravitational wave frequency bin, $d(\textnormal{ln}\fgearth)$. This is done first by converting from co-moving volume shell to redshift, and then converting from redshift to Earth-observed time.

Given Eqs.\,\ref{eq:hrms}-\ref{eq:d4N}, $\hc$ can be written as a power-law with dimensionless amplitude $\ayr$ \citep{jenet+06}:
\begin{equation}\label{eq:hpowerlaw}
\hc(f) = \ayr \left( \frac{f}{{\rm yr}^{-1}} \right)^{-2/3}~.
\end{equation}
PTA constraints are often quoted as an upper limit on $\ayr$ using this basic $f^{-2/3}$ model of the expected SMBH binary background. 
However, PTAs are most sensitive to the GW frequency close to \mbox{$f\rightarrow1/T_{\rm obs}$}, where $T_{\rm obs}$ is the total observation time, which for current data sets is around ten years. The constraint on $\ayr$ is extrapolated using a spectral model of $f^{-2/3}$. A super-efficient inspiral from environmental coupling could give rise to a different spectral shape of the GWB \citep{sesana13CQG,ravi+14,huerta+15}, which may include a ``turn-over'' frequency in the PTA band, where the binary evolution changes from being dominated by energy removal from environmental coupling to being dominated by gravitational radiation \citep{sampson+15}.
In this paper, we work within the single-power-law model, but as PTA limits continue to improve, the extrapolation to a meaningful limit on $\ayr$ will become increasingly dependent on the assumed shape of the spectrum \citep{nanograv+15}.

\subsection{Parameters of Galaxy Evolution}
There are no current direct observational constraints on the demographics of supermassive black hole (SMBH) binaries. This work follows the prescription used in \citet{sesana13} and \citet{ravi+15}, where the galaxy merger rate is used as a proxy and each galaxy in a merger is populated with a central SMBH using the observed relationship between SMBHs and their host galaxies.

The number density of galaxy mergers
\begin{equation}
 \frac{d^{3}n_{\rm G}}{dz ~dM ~dq} = \Phi (z, M) ~\mathcal{R} (z, M, q)~,
\end{equation}
is a combination of two functions: the galaxy stellar mass function (GSMF), $\Phi (z, M)$, and the galaxy merger rate $\mathcal{R} (z, M, q)$.


The GSMF is an astronomical observable, and is typically parametrized in terms of a either a single or a double Schechter function:
\begin{equation}
  \begin{aligned}
    \Phi (z, M) &= \left. \frac{dn_{\rm G}}{dM} \right\vert_{z} \\
    &= \frac{e^{-\frac{M}{M^{\ast}}}}{M^{\ast}} \left[ \Phi^{\ast}_{1} \left( \frac{M}{M^{\ast}} \right)^{\gamma_{1}} + \Phi^{\ast}_{2} \left( \frac{M}{M^{\ast}} \right)^{\gamma_{2}} \right]~.
  \end{aligned}
  \label{eqn:gsmf}
\end{equation}

The galaxy merger rate is the redshift-dependent rate at which a galaxy of mass $M$ is involved in a major merger ($q > \frac{1}{4}$) with a second galaxy with mass $qM$:
\begin{equation}
 \mathcal{R} (z, M, q) = \frac{d^{2}n_{\rm G}}{dz ~dq} = \frac{df_{\rm pair} (z)}{dq} \frac{1}{\tau (z, M)} \frac{dt}{dz}~.
 \label{eqn:GMRate}
\end{equation}
This is calculated by combining the galaxy pair fraction, $\fpair(z)$, and the merger timescale, $\tau (z, M)$. The galaxy pair fraction is an astronomical observable. The merger timescale, which approximates the dynamical friction timescale for a dynamically bound pair of galaxies, must be determined through simulations.

Once the number of galaxy mergers has been calculated in a given redshift, mass and mass ratio range, the mass of each black hole in the merger is calculated using SMBH-galaxy scaling relations that have been observed in the local Universe. Again, we use the prescription described in \citet{sesana13} to calculate $\mbulge$ for each galaxy assuming some fraction of the total mass is in the central bulge. This fraction, $\fbulge$, depends on galaxy morphology and mass. The SMBH mass is then calculated using a relationship of the form:
\begin{equation}
    {\rm log}_{10}M_{\bullet} = \alpha + \beta \,{\rm log}_{10}\bigg(\frac{\mbulge}{10^{11}\msun}\bigg)~.
    \label{eqn:MSig}
\end{equation}
Additionally there is an intrinsic scatter, $\epsilon$, associated with each measurement of this relation, which is the natural scatter of individual galaxies around the trend line shown above.

Combining Equations \ref{eq:strain} - \ref{eqn:MSig}, this model gives a prediction for $\ayr$:
\begin{equation}
    \ayr^{2} = \int \int \int ~\frac{\Phi}{\tau} \frac{df_{\rm pair}}{dq} \frac{dV_{c}}{dz} \left. \left( \frac{dt}{d\left( {\rm ln} f \right)} h_{s}^{2} \right) \right\vert_{f_{\rm yr}}~dz~dM~dq.
    \label{eqn:a^2}
\end{equation}


\section{Observational Constraints}\label{sec:obscon}
\citet{sesana13} presented the first systematic investigation of the GWB using the method described in Section \ref{sec:model} and utilized the mean values of many different observations. This work seeks to update that investigation in a complimentary way to what was done in \citet{ravi+15}, which used fewer, but more complete observations and focused closely on the uncertainties in the galaxy merger rate as well as other uncertainties inherent to the specific observations used to determine the GSMF and the $\mmb$ relation. This work attempts to find a balance between these two approaches, using multiple current and complete observations, while also attempting to parse which details are more relevant to $\ayr$ predictions.

\subsection{Galaxy Stellar Mass Function}
The total GSMF is fairly well-constrained by observation out to $z = 1.5$ \citep{tomczak+14}, however there is still debate about how the total GSMF is broken down by galaxy type (star-forming vs. quiescent). Quiescent galaxies dominate the GW radiation that PTAs are sensitive to because they host larger black holes than star-forming galaxies. Therefore, the break-down of galaxy type as a function of mass and redshift has a large impact on the overall signal. We do not assume any kind clustering effect on mergers, so a quiescent galaxy's likelihood of merging with another quiescent is only based on the fraction of quiescent galaxies in a given redshift, mass, and mass ratio range.
\citet{ravi+15} shows that while there is some possibility for contamination of colour-selected GSMFs, the overall effect on the amplitude of the GWB is negligible compared to other contributions.

In this paper, we use the GSMF measured by the CANDELS/ZFOURGE survey \citep{tomczak+14}, and the full UltraVISTA survey \citep{ilbert+13} for $z > 0.2$. The ZFOURGE survey provides the deepest GSMF measurements to date \citep{tomczak+14}, while the UltraVISTA survey covers a wide field and so they compliment each other well.
These two surveys show consistent results with the updated GSMF from \citet{muzzin+13} used by \citet{ravi+15}. This paper uses the SDSS/GALEX observations of the local universe GSMF ($z < 0.2$; \citealt{moustakas+13}).

\subsection{Galaxy Merger Rate \& Timescale Between Mergers}\label{subsec:gmr}
The observed galaxy merger rate is a combination of the observed galaxy pair fraction $\fpair$, and an analytically calculated merger timescale $\tau$ (Eqn.~\ref{eqn:GMRate}). The galaxy pair fraction, $\fpair$ is fairly well constrained in the local universe ($z \sim 0.1$), but it is less constrained with increasing redshift \citep{keenan+14}. $\fpair$ is expected to peak and turn-over somewhere around $z \sim 2-3$ for the most massive galaxies, which we are interested in \citep{conselice14}. In this paper, we use the galaxy pair fraction from the GAMA survey \citep{robotham+14} and the combined value resulting from analysis of the RCS1, UKIDSS, and 2MASS surveys \citep{keenan+14}. While these two papers agree on the pair fraction in the local universe ($z \sim 0.1$), they differ by a factor of two on the redshift dependence of the pair fraction, with \citet{keenan+14} showing a lower value. Here we keep the values of $f_{\rm pair}$ fixed out to $z = 3$, and find that the values of $f_{\rm pair}$ reached are approximately at both the high and low ends of predictions, respectively \citep{conselice14}. While both \citet{sesana13} and \citet{ravi+15} use the $\fpair$ measurement from \citet{xu+12}, it is not used in this work because the authors of \citet{robotham+14} note that a local pair fraction observation from \citet{xu+12}, which sets both the low fraction at $z=0$ and the high slope of the redshift dependence of the presented $\fpair$ relation, has been shown to be unrealistic. The revised measurement of \citet{xu+12} is shown to be consistent with \citet{robotham+14}. 

For self-consistent comparisons, recent observations of the galaxy pair fraction use one of two formulas for $\tau$ \citep{lotz+11,kw08}, which differ by a factor of two. The \citet{kw08} timescale is a lower limit for the galaxy merger timescale because it is derived by a dark matter halo merger timescale from the Millennium Simulation. \citet{lotz+11} calculates $\tau$ from a set of hydrodynamical simulations, which incorporates gas and dust into the galaxy merger, unlike \citet{kw08}. \citet{xu+12} combined the mass and redshift dependence from \citet{kw08} with the results from \citet{lotz+11} to give a description of the major-merger timescale for $z < 1$
which is combined with the pair fractions described above to give two different galaxy merger rates used in this work.
Since this paper is focused on the effects of observable galaxy evolution parameters on the GWB we do not investigate different formulations of tau. However, in \citet{ravi+15} the removal of the mass and redshift dependence of $\tau$ was found to produce a slightly higher value for $\ayr$.

\begin{figure}
    \centering
    \includegraphics[trim=15mm 28mm 10mm 25mm,clip,width=\columnwidth]{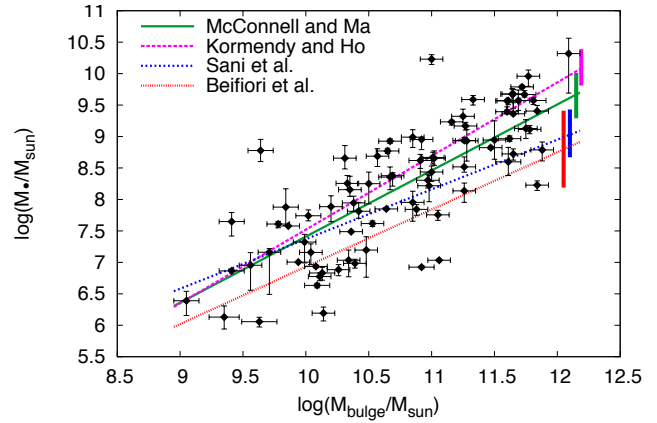}
    \caption{Demonstration of the variation in $M_\bullet$-$M_{\rm bulge}$ correlations in the literature. The intrinsic scatter parameter, $\epsilon$, is shown as a $\pm$ vertical bar for each relation. The points are given by the measurements in the review of \citet{kormendyho13}. The differences in $\alpha$ and $\beta$ measurements depend on a number of effects including: 1) mass range considered; 2) inclusion of pseudobulges/other subsample effects; and 3) how bulge mass is determined (dynamical vs.\ stellar luminosity). The major difference between the KH13/MM13 fit and the \citet{sani+11}/\citet{beifiori+12} fits is the inclusion of new measurements of galaxies at the highest-mass end for KH13/MM13.
    }
    \label{fig:mmbulge}
\end{figure}

\subsection{SMBH mass - Galaxy Bulge Mass Relation} \label{sec:GBMR}
In the past few years, there have been observations of a steeper relationship than previously observed for the correlation between bulge mass and the mass of the resident SMBH. The steeper values are largely due to the recent measurements of the high-mass end of $M_{\rm bulge}$, which in turn house the most massive SMBHs that will contribute to the GW signal in the PTA band \citep{scott+13,kormendyho13,mcconnellma13}. As we aim to predict the most realistic GWB signal as relevant to PTAs, in our simulation we thus only consider the $M_{\bullet}$-$M_{\rm bulge}$ relations that include these massive galaxy measurements. Figure \ref{fig:mmbulge} shows a demonstration of how the parameters of Eq.\,\ref{eqn:MSig} vary with the inclusion or exclusion of measurements at the high-mass end, and various other considerations. It is unclear whether a single power-law best describes this relationship over the full mass range \citep{graham16}, but it appears to be sufficient for the most massive systems, and so in this work we do not consider the broken power-law prescriptions in \citet{scott+13}. Additionally, we note that the high mass relation in \citet{scott+13} is almost identical to the relation in \citet{kormendyho13}.

Measurements of the $\mmb$ relation also include $\epsilon$, the ``intrinsic scatter'', which is the natural scatter of individual galaxies around the trend line described by $\alpha$ and $\beta$ in Eqn.\,\ref{eqn:MSig}. This parameter plays a critical role in $\ayr$ predictions as a way accounting for the outliers in the distribution. The galaxies containing over-massive black holes will contribute more to the GW signal in the PTA band than most galaxies of similar mass and thus are a necessary inclusion into any prediction of $\ayr$.

\begin{figure}
    \centering
    \includegraphics[width=\columnwidth]{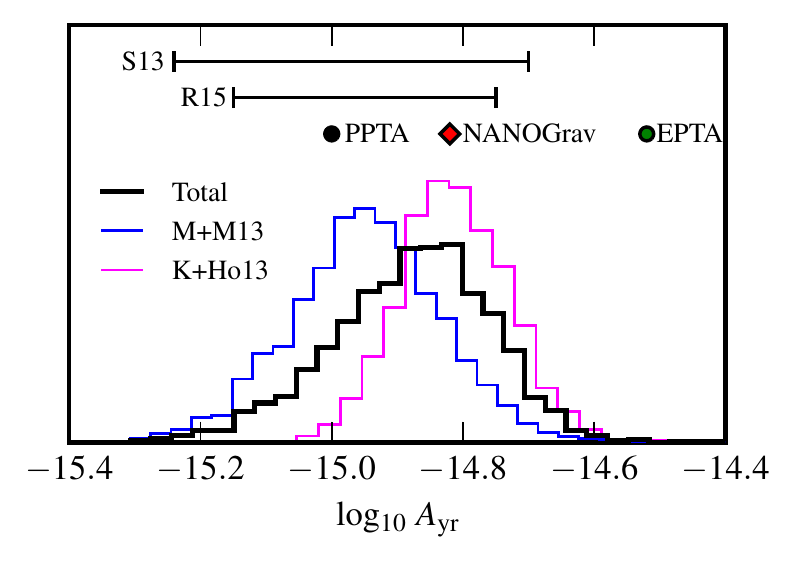}
    \caption{Predictions of $\ayr$ calculated using Eqn.\ref{eqn:a^2}. The solid black line is the total distribution calculated using the errors reported with the measurements we have chosen to use, described in \S\ref{sec:obscon}. Additionally, the total distribution is broken down by $\mmb$ relation, with \citet{kormendyho13} shown in magenta, which is understood to predict higher values of $\ayr$ because the measurement is weighted strongly by large mass observations, as seen in Fig\ref{fig:mmbulge}, while \citet{mcconnellma13} shown in blue is not, and thus creates smaller black hole masses. The breadth of the distributions results from the observed variance in our input parameters, and also from the fact that we use multiple formulations for the GSMF and $f_{\rm pair}$. The most recent upper limits from each PTA are shown as points directly above the distribution of $\ayr$, and above that we show the predictions from similar models in \citet{sesana13} and \citet{ravi+15}.
    }
    \label{fig:Ayr}
\end{figure}


\section{Results}\label{sec:results}
Combining all of the observational constraints from \S\ref{sec:obscon} into our model described in Eqn. \ref{eqn:a^2}, we calculate a range of predictions for $\ayr$. The inputs are allowed to vary within the reported observational errors for each combination of GSMF, $\fpair$, and $\mmb$ relation used in this paper. Each combination is run $500$ times for a total of $4000$ predictions of $\ayr$.
Figure \ref{fig:Ayr} shows a plot of this distribution for major-mergers, $q > \frac{1}{4}$, between galaxies with a stellar mass in the range $10^{10} - 10^{12} \msun$ and lying in a redshift range $z < 3$. Additionally, we include the predictions from \citet{sesana13} and \citet{ravi+15}, and find that all ranges on $\ayr$ are similar. Fig \ref{fig:Ayr} also indicates where recent PTA upper limits fall in respect to this model's predictions. The most recently published upper limit comes from 11 years of data taken by the Parkes Pulsar Timing Array \citep[PPTA][]{ppta+15}, which quotes an upper limit on $\ayr$ of $1.0\times10^{-15}$. The North American Nanohertz Observatory for Gravitational Waves (NANOGrav) has released nine years of data which produce an upper limit of $\ayr<1.5\times10^{-15}$ \citep{nanograv+15}, and the European Pulsar Timing Array (EPTA) quoted an upper limit of $\ayr$ of $3.0\times10^{-15}$ \citep{EPTA+15}. PTA upper limits are typically shown to ``rule out'' some portion of the predicted range on $A_{\rm yr}$. However, it is often unclear what that means in terms of limits on the input SMBH evolution parameters. We aim to provide clarity by determining how, and how much, each parameter effects the resulting prediction of $A_{\rm yr}$.

\begin{figure}
    \centering
    \includegraphics[width=\columnwidth]{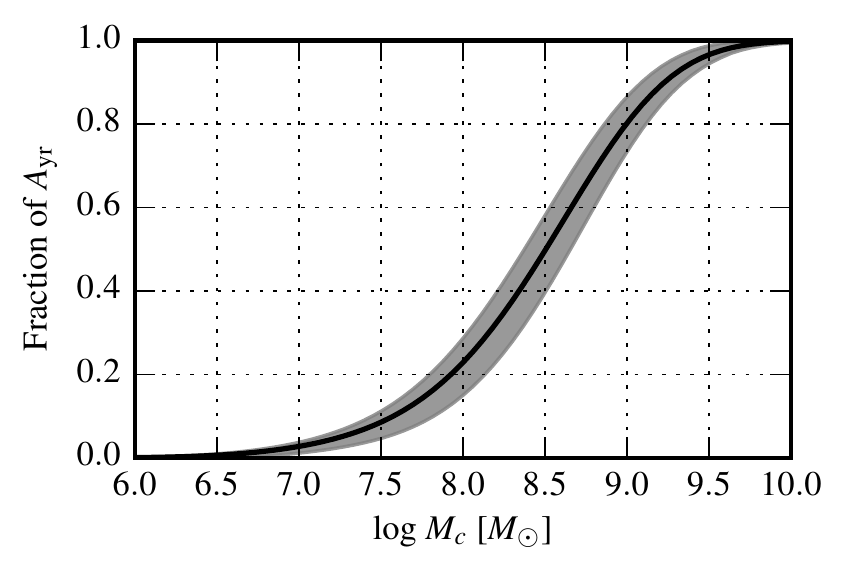}
    \caption{Cumulative fractional contribution to $\ayr$ from binaries over a range of chirp mass. Each prediction of $\ayr$ is the quadrature sum of many binaries. The solid black line shows the cumulative fraction of $\ayr$ that comes from binaries with a given $M_{c}$. The gray region shows the one sigma error region for this distribution. The value of $\ayr$ is dominated by binaries with $M_{c} > 10^{8} \msun$, as expected.}
    \label{fig:McvsAyr}
\end{figure}

\begin{figure}
    \centering
    \includegraphics[width=\columnwidth]{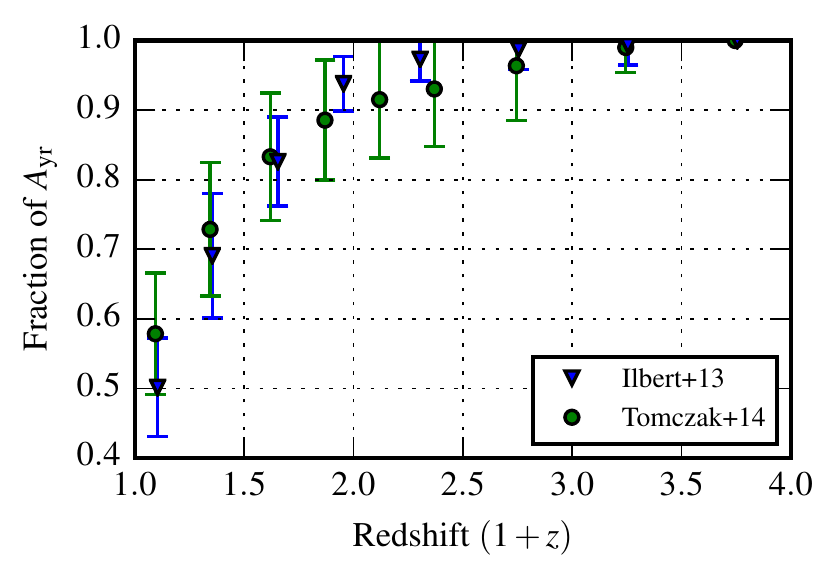}
    \caption{Cumulative fractional contribution to $\ayr^2$ as a function of redshift for the two GSMFs used in this work. The redshift distribution of $\ayr$ of the GSMF from \citet{ilbert+13} is plotted as blue triangle points, while the GSMF from \citet{tomczak+14} is plotted as the green circular points.The GSMF provides the mass distribution of galaxies as a function of redshift, and thus affects how each redshift bin will contribute to the total value $\ayr$. Above we plot both GSMFs with error bars, which indicate the root-mean-squared value from all simulation runs.
    While the majority of the contribution to $\ayr$ comes from binaries at $z \lesssim 1.5$, there is a large amount of variance which is dominated by the uncertainty in the amount of massive galaxies in this redshift range. We discuss the relative contributions from variance in the GSMFs on the variance of $\ayr$ predictions in Section\,\ref{sec:error}.
    }
    \label{fig:ZvsAyr_gsmf}
\end{figure}

\begin{figure}
    \centering
    \includegraphics[width=\columnwidth]{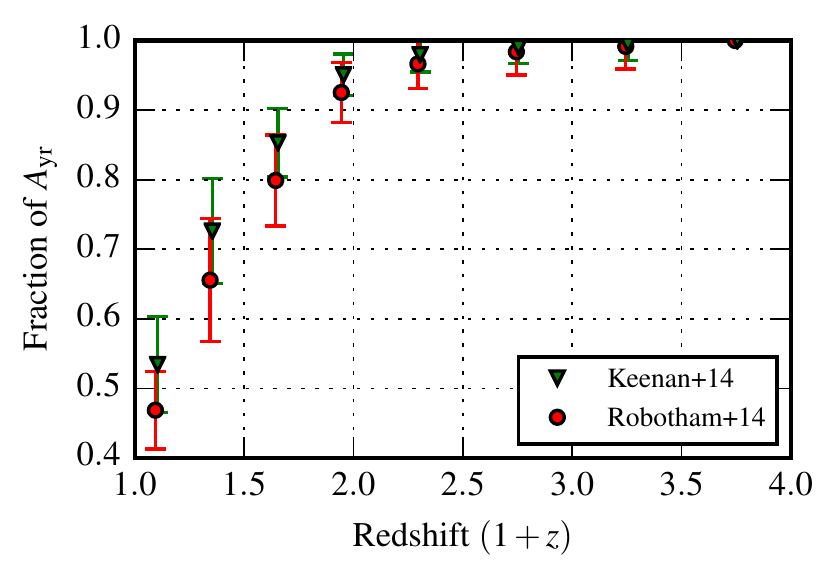}
    \caption{Cumulative fractional contribution to $\ayr^2$ as a function of redshift for the two different observational measurements of $f_{\rm pair}$ used in this work. $f_{\rm pair}$ is the fraction of galaxy pairs in a certain mass, redshift and mass ratio range that are undergoing a merger. This parameter sets the number of sources that will be contributing to $\ayr$. While both measurements have the same value at $z = 0$, the redshift dependencies vary by a factor of two. The higher slope $f_{\rm pair}$ from \citet{robotham+14}, shown above in red circles, contributes less at lower redshifts then the lower slope $f_{\rm pair}$ from \citet{keenan+14}, shown above in green triangles, which is directly related to the number of binaries contributing to $\ayr$ at higher redshifts.
    }
    \label{fig:ZvsAyr_fpair}
\end{figure}

As expected for each prediction, $\ayr$ is dominated by binary sources with chirp masses larger then $10^{8}\,\msun$ as seen in Figure \ref{fig:McvsAyr}, which shows how the cumulative contribution to $\ayr$ from increasing chirp masses. Figures \ref{fig:ZvsAyr_gsmf} \& \ref{fig:ZvsAyr_fpair} show that the majority of the signal is produced at $z \lesssim 1.5$, also as expected, but there is a lot of variance at lower redshift, which is discussed further Section \ref{sec:error}. Figure \ref{fig:ZvsAyr_gsmf} shows the contribution from binaries at different redshifts for both GSMFs used in this paper. While the general trend is the same for each, the differences in them directly follow the way that each GSMF handles the abundance of massive early-type galaxies, specifically between $0.5 < z < 1.5$.

At higher redshifts the amount of signal is dominated by the total number of binaries, which is set in part by $f_{\rm pair}$. The effect of different $f_{\rm pair}$ on the redshift distribution of the contribution to $\ayr$ is shown in Figure \ref{fig:ZvsAyr_fpair}. While both values of $f_{\rm pair}$ used in this paper have the same value at $z = 0$, the redshift dependencies vary by a factor of two. However, as described in Section \ref{subsec:gmr}, the two values trace the upper and lower ends of predictions for $\fpair$ at high redshift. An interesting feature of Figure \ref{fig:ZvsAyr_fpair} is that the higher slope $f_{\rm pair}$ from \citet{robotham+14} contributes less at lower redshifts than the the lower slope $f_{\rm pair}$ from \citet{keenan+14}. This trend is directly related to the number of binaries contributing to $\ayr$ at higher redshifts.

\subsection{Relative effects of GSMF, $\fpair$, and $\mmb$} 
\label{sec:error}
\begin{figure}
    \centering
    \includegraphics[width=\columnwidth]{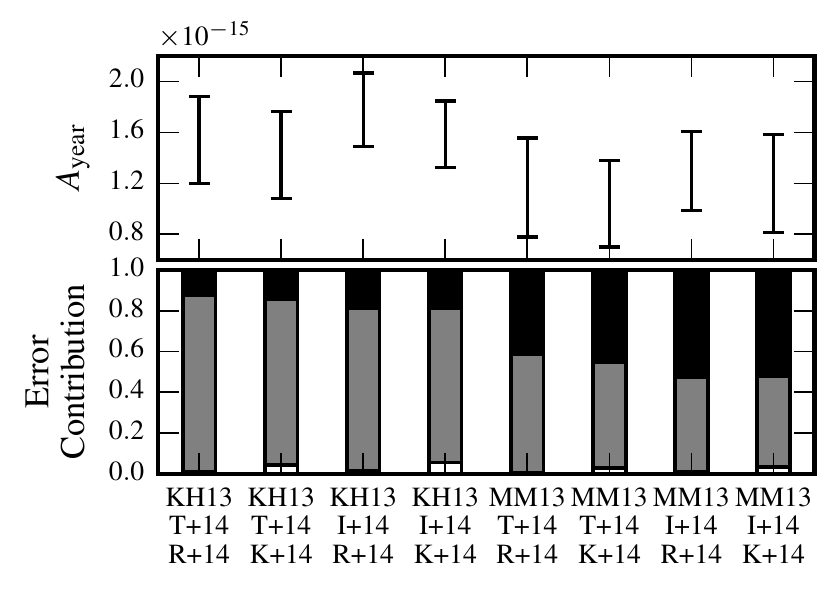}
    \caption{Error contribution to $\ayr$ from observations of galaxy evolution parameters.
    In the top portion of this graph, the error bars show the one sigma region of predictions for $\ayr$ for each combination of galaxy evolution parameters used in the model. The bottom portion of the graph shows the stacked error contributions from each parameter. The black region is from the $\mmb$ relation, while the gray region is from the GSMF, and the white region is from $\fpair$. The label on each combination shows which measurement was used: KH13 - \citep{kormendyho13}, MM13 - \citep{mcconnellma13}, T+14 - \citep{tomczak+14}, I+14 - \citep{ilbert+13}, R+14 - \citep{robotham+14}, K+14 - \citep{keenan+14}. Clearly the error from the GSMF measurements dominate the error budget of each prediction, followed by the error from the $\mmb$ relation. The errors in $\fpair$ are minimal as there is barely a hint of white in each stacker bar graph. The implications of this plot are discussed in \S\ref{sec:error}.
    }
    \label{fig:Table1}
\end{figure}

Each parameter in Eqn.\ref{eqn:a^2} that has an observational constraint also has an associated error. These errors obviously effect the variance in the final strain spectrum prediction. In Fig. \ref{fig:Table1}, we show the $1-\sigma$ range of predictions for $\ayr$ given by each combination of observational parameters. The percentage contribution to the total error is broken down for each parameter in a given combination, and shown underneath the predicted range of $\ayr$. Below, we discuss the breakdown of the errors in \emph{individual observational parameters} impact on the final range of $\ayr$.

Given that $\ayr^{2} \propto M_{c}^{5/3}$, the parameters which affect the binary chirp mass should have substantial impact on the prediction of $\ayr^{2}$ \citep{sesana10CQG}. Accordingly, we find the $\mmb$ relation sets the mean value for $\ayr^{2}$ predictions, as can be seen by the breakdown in Figure \ref{fig:Ayr}. This is the relation that provides the translation from galaxy population to BH mass and therefore sets the mean value of the $M_{c}$ distribution. Section \ref{sec:astrolimsmbulge} discusses in detail how the $\mmb$ measurement parameters contribute to $\ayr^2$.

The GSMF plays a large role in determining the variance of a certain prediction. All mass functions used in this paper are parametrized using a double Schechter function (Eqn \ref{eqn:gsmf}) for $z < 1.5$, so there are a lot observed parameters included in each calculation. Looking at each Schechter parameter and the error bars associated with it reveals that uncertainty in $M^{\ast}$ has the largest effect $\ayr^2$ predictions, contributing $45\%$ of the GSMF error for both observations used in this paper. $M^{\ast}$ is the mass at which the Schechter function transitions between the high-mass exponential decay and the lower-mass slopes $\alpha_{1}$ and $\alpha_{2}$. The error from the local GSMF, $z < 0.2$ \citep{moustakas+13}, is minimal at $2\%$ of the GSMF error, but the next redshift bin, $0.2 < z < 0.5$, in both GSMF observations used in this paper provides the largest influence by redshift on GSMF error contribution to $\ayr^2$ at $55\%$ for \citet{tomczak+14} and $72\%$ for \citet{ilbert+13}.

\begin{figure}
    \centering
    \includegraphics[width=\columnwidth]{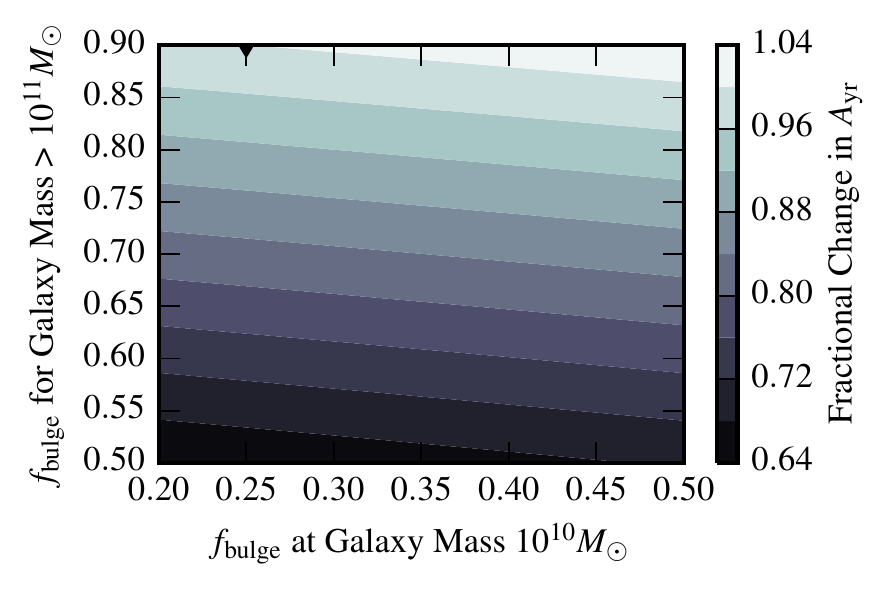}
    \caption{Influence of $\fbulge$ on $\ayr$. $\fbulge$ describes the fraction of a galaxy's total mass contained in the central bulge. This parameter is used to estimate a black hole's mass using the host galaxy's total stellar mass. For quiescent galaxies there are two parameters used to set $\fbulge$, a value for massive galaxies, $M > 10^{11} \msun$, which is plotted on the y-axis, and a value for the less massive galaxies, $M = 10^{10} \msun$, which is plotted on the x-axis. For galaxies with masses between $10^{10} - 10^{11} \msun$, $\fbulge$ is assumed to follow a log-linear line between the set values. Both of these values were allowed to vary and the resulting fractional change in $\ayr$ is shown on the z-axis in color. In this paper we use the values (0.25, 0.9), which is the point used for comparison in the fractional change shown in the color bar. Overall there is not a strong dependence on $\ayr$ from the parameterization of $\fbulge$.
    }
    \label{fig:fbulge}
\end{figure}

The host galaxy's mass is used to estimate each black hole's mass using $\fbulge$. $\fbulge$ for quiescent galaxies is described with two values, one that sets the fraction of mass in the central bulge for massive galaxies, $M > 10^{11} \msun$, and one that sets the fraction of mass in the central bulge for galaxies with stellar masses of $10^{10} \msun$. Given the dominance of the signal from binaries with chirp masses above $10^{8} \msun$, the effect of the quiescent galaxy's $\fbulge$ on $\ayr$ was calculated. Fig.\,\ref{fig:fbulge} demonstrates the dependence of $\ayr$ on $\fbulge$ for quiescent galaxies, making it clear that there is not a strong dependence (less than a factor of $\sim\sqrt{2}$).

The pair fraction, $f_{\rm pair}$, has the least impact on $\ayr^2$. Although the parameter itself has a large range, it only impacts the number of sources ($\ayr^2 \propto N$), rather than the masses of those sources. 

\subsection{Translating GWB Limits and Astrophysical Parameters}\label{sec:trans}
The overarching goal of this paper is to provide a mapping of GWB limit/measurement values to specific parameters of galaxy formation. Below, we look at how areas of parameter space correspond to specific values of $\ayr$ with an associated error range. In our discussions below we aim to make the following two statements attainable:
\begin{itemize}
\item Given a PTA upper limit on the GWB, what specifically does this mean for galaxy/SMBH evolution?
\item Given a new observation of the \mmb\ relation, are the new values compatible with the best PTA limit? If not, how can they be reconciled?
\end{itemize}

\subsubsection{GW limits and the Black Hole - Host Galaxy Relation}\label{sec:astrolimsmbulge}
\begin{figure}
    \centering
    \includegraphics[width=\columnwidth]{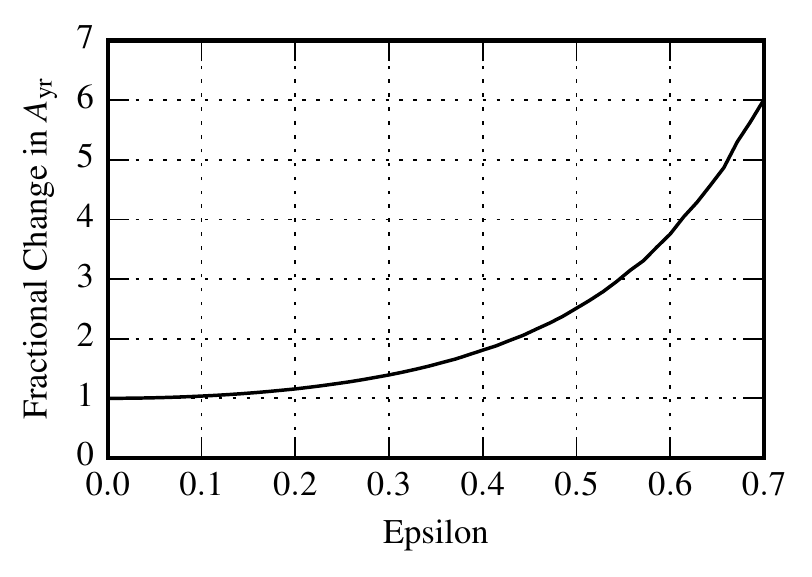}
    \caption{Relative scaling of $\ayr$ as a function of the intrinsic scatter, $\epsilon$, of the $\mmb$ relation. As the scatter increases, the black hole mass function adjusts to include higher mass SMBH systems and as such the overall strain increases in an exponential manner.
    }
    \label{fig:hvse}
\end{figure}

The parameter space that characterizes the black hole-host galaxy relation is encompassed by $\alpha$, $\beta$, and $\epsilon$. The last of these is the natural scatter of individual galaxies around the trend described by intercept $\alpha$ and slope $\beta$. This $\epsilon$ plays a critical role in $\ayr$ predictions; as $\epsilon$ increases, higher masses are present in a few binaries, which increases the distribution of chirp masses and effectively weights $\ayr$ upward.
We demonstrate the implications of this in Figure \ref{fig:hvse}, which shows the dependence calculated numerically; for any combination of $\alpha$ and $\beta$, changes in $\epsilon$ effect $\ayr^2$ in the same way.

\begin{figure}
    \centering
    \includegraphics[width=\columnwidth]{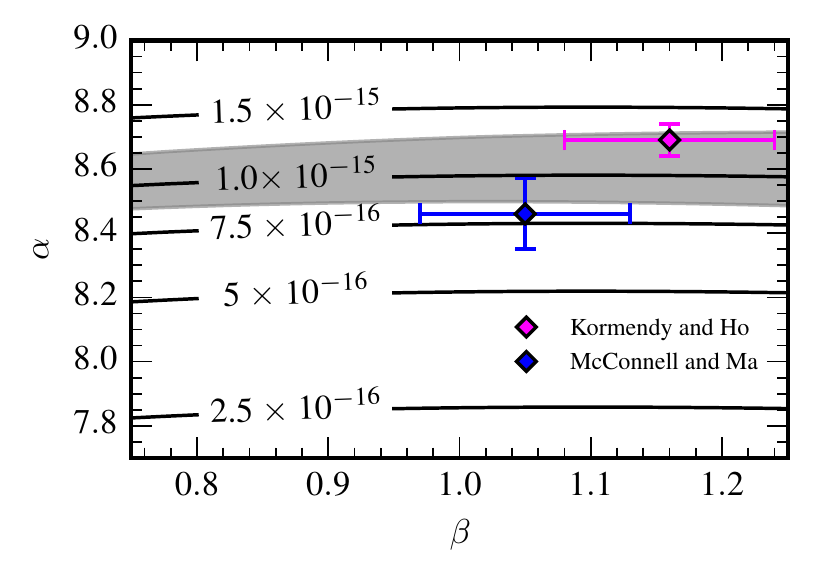}
    \caption{Mapping of GWB amplitude to $\mmb$ parameters $\alpha$ and $\beta$. These parameters, with $\epsilon$, are used to characterize the black hole-host galaxy relation described in Eqn.\,\ref{eqn:MSig} ($\epsilon=0$ in this figure). The contours show constant values of $\ayr$; note, however, that the contours will scale downward with $\epsilon>0$ according to the curve in Fig.\,\ref{fig:hvse}. Observational measurements are indicated by diamonds. The grey region accounts for uncertainties in other measured astrophysical parameters, specifically the GSMF and the galaxy merger rate. This uncertainty applies to all values of $\ayr$ but is only shown on the current best published PTA limit. A set value of $\ayr$ translates to a collection of $\alpha-\beta$ pairs, while a PTA upper limit on the value of $\ayr$ translates to an upper limit on these parameters. Values of $\alpha$ and $\beta$ which predict a larger $\ayr$ would be inconsistent with that PTA limit.
    }
    \label{fig:avsb}
\end{figure}

\begin{figure}
    \centering
    \includegraphics[width=\columnwidth]{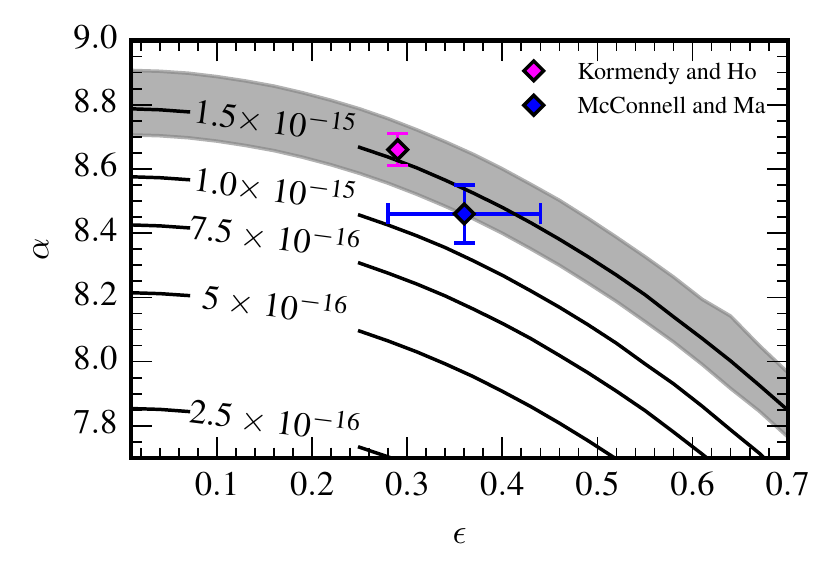}
    \caption{Mapping of GWB amplitude to $\mmb$ parameters $\alpha$ and $\epsilon$. These parameters are found to most strongly define the predicted strength of GW amplitude. The contours show the value of the $\ayr$ as a function of these parameters. Observational measurements are shown as indicated. The grey region accounts for uncertainties in other measured astrophysical parameters, specifically the GSMF and the galaxy merger rate. As in Fig.\,\ref{fig:avsb}, this uncertainty applies to all values of $\ayr$ but is only shown on the current best published PTA limit. A PTA upper limit on the value of $\ayr$ translates to an upper limit on the parameters $\alpha$ and $\epsilon$ and would be inconsistent with parameter values that predict a larger $\ayr$.
    }
    \label{fig:avse}
\end{figure}

In Fig.~\ref{fig:avsb}, we show contours of constant $\ayr$ in $\alpha$-$\beta$ space, while keeping $\epsilon$ equal to zero, as this is the most common way that the $\mmb$ relation is reported. Yet, for a specific value of $\alpha$, changing $\beta$ across the entire range of observed values translates to only a $<20\%$ change in $\ayr$.
It is thus $\alpha$ and $\epsilon$ that have the most impact on a prediction of $\ayr$, and so we combine the results of Fig.~\ref{fig:hvse} \& \ref{fig:avsb} in Fig.~\ref{fig:avse} to show $\ayr$ as a function of $\alpha$ and $\epsilon$, while holding $\beta=1$.

\subsubsection{GW limits and Stalling Binaries}
\begin{figure}
    \centering
    \includegraphics[width=\columnwidth]{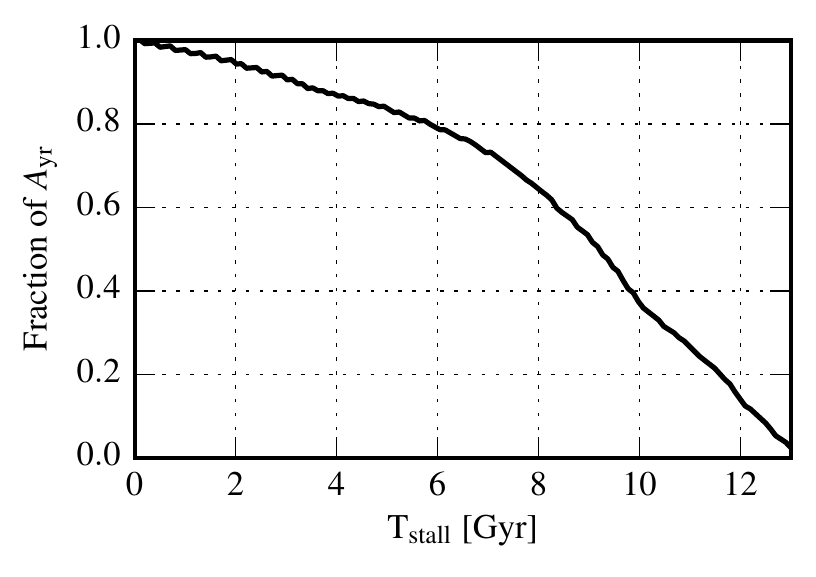}
    \caption{Effect of SMBH binary stalling time on the expected amplitude of the GWB. $T_{\rm stall}$ measures the time between the galaxy merger and the binary SMBH entering the PTA band in Gyrs. The predicted value of characteristic strain for a given value of astronomical parameters depends on the value of $T_{\rm stall}$. As the time increases, binaries ``stall'' for longer and less reach the PTA band, which lowers the overall strain until the time approaches a Hubble time and there is effectively no GWB from binary SMBHs.
    }
    \label{fig:hvsTstall}
\end{figure}

The assumption that galaxy mergers and binary black holes form at the same cosmological time has until now been implicit in the model used in this paper and others. Yet, as PTA upper limits become inconsistent with measured astronomical parameters, the assumptions of this model must be questioned. It is straight forward to ease this assumption by allowing for a ``stall'' in the binary SMBH formation. Let us introduce a variable, $T_{\rm stall}$, which is a measure of the time between the galaxy merger, which occurs approximately on a dynamical friction timescale, and the binary SMBH entering the PTA band.
This stalling timescale creates a redshift offset between the galaxy merger and binary SMBH's in-band GW emission. This is incorporated into Eqn.\,\ref{eq:d4N} like so:
\begin{equation}
\frac{d^{4}N}{dz ~dM ~dq ~d(\textnormal{ln}f_{r})}  = \frac{d^{3}n}{dz_{1} ~dM ~dq} ~ \frac{dV_{c}}{dz_{2}} ~ \frac{dz_{2}}{dt_{r}}	~ \frac{dt_{r}}{d(\textnormal{ln}f_{r})}~,
\end{equation}
where $z_{1}$ is the redshift of the galaxy merger, and $z_{2}$ is the redshift at which the binary is emitting in the PTA band, while $T_{\rm stall}$ is the proper time between $z_{1}$ and $z_{2}$.
Fig.~\ref{fig:hvsTstall} shows how $\ayr$ changes with different values of $T_{\rm stall}$. Obviously as the stalling time scale reaches values nearing the Hubble time, $\ayr$ falls to zero as no systems are expected to ever enter the PTA band. The meaning and use of limits on $\tstall$ are described further in Section~\ref{sec:stall}.

\section{Limiting $\mmb$ and stalling timescale with PTA constraints}\label{sec:ptacon}
The following section describes a method by which PTA constraints can be extended beyond $\ayr$ and into the parameter space of galaxy evolution. Similar goals have been proposed using different methods in \citet{middleton+16}. We note that a key difference between this work and that of \citet{middleton+16} is that we interpret PTA limits in the context of other observational constraints on galaxy evolution, whereas the \citet{middleton+16} method assumes no outside knowledge of the SMBH binary population, except for a general form of the SMBH merger rate density (which is well motivated), to formulate their constraints. As such, they were unable to place meaningful limits on the binary population, while this work is able to place tighter constraints by utilizing a well established range of prior information.

\subsection{Constraining $\alpha$, $\beta$, and $\epsilon$ with PTAs}
It is common for PTA constraints to be quoted as a single number, $\ayr$, which represents the $95\%$ upper limit on a GWB of a given spectral index: for binary SMBHs, $\alphah=-2/3$ (Eq.\,\ref{eq:hpowerlaw}). However, quoting a single number is only for simplicity; the actual result produced by PTAs for a limit on a power-law GWB is a probability distribution for the value of the strain amplitude.

Here we describe the use of this probability distribution to obtain direct limits on $\alpha$, $\beta$, and $\epsilon$. We then demonstrate how an inconsistency with an observed value can be used to place a lower limit on $\tstall$. 

\begin{figure}
    \centering
    \includegraphics[trim=2mm 2mm 2mm 2mm,clip,width=\columnwidth]{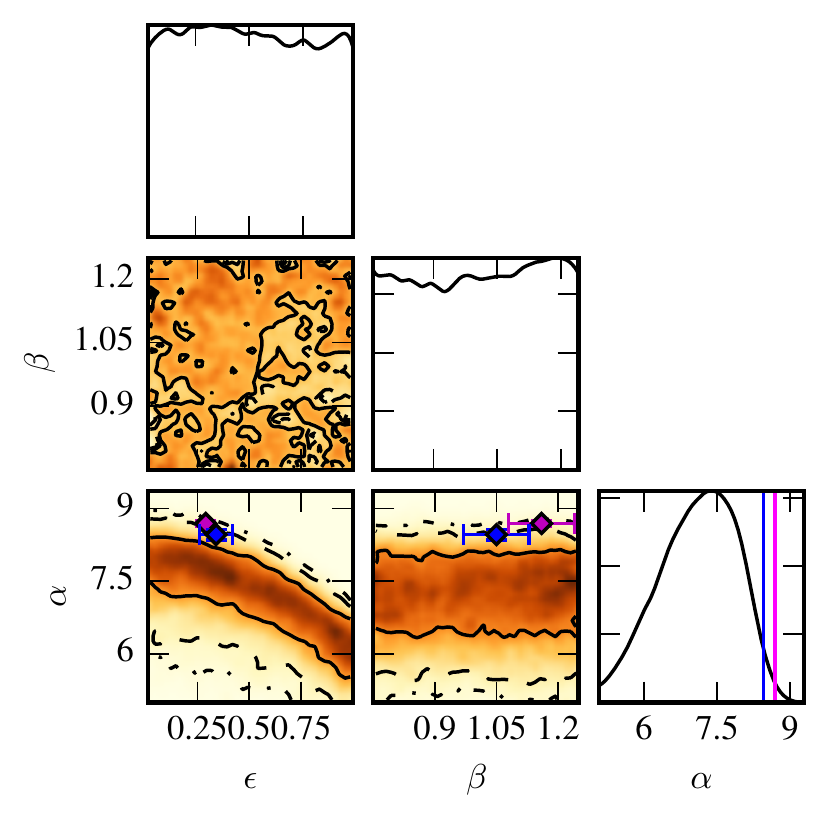}
    \caption{Translation of the marginalized posterior distribution of $\ayr$ into the black hole-host galaxy parameter space, which is characterized by an intercept $\alpha$, a slope $\beta$, and an intrinsic scatter $\epsilon$. $\beta$ is not informed by the distribution of $A_{\rm gw}$, while both $\alpha$ and $\epsilon$ are, with a limit on $\alpha$ being more strongly set. The curves show the 1, 2, and 3$\sigma$ contours. Since $\beta$ is not strongly informed by the upper limit, we can set an upper limit in $\alpha$-$\epsilon$ space by marginalizing over $\beta$. This upper limit and relevant observational measurements of these parameters are shown in Fig.~\ref{fig:Bayes_avse}.
    }
    \label{fig:TriPlot_abe}
\end{figure}
\begin{figure}
    \centering
    \includegraphics[width=\columnwidth]{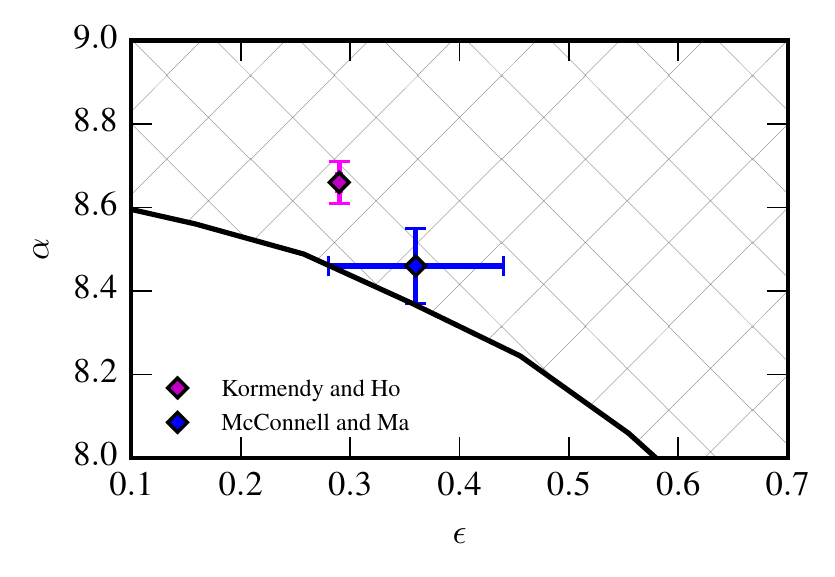}
    \caption{Translation of the $95\%$ upper limit on $\ayr$ into the parameter space $\alpha$-$\epsilon$, which characterizes the black hole-host galaxy relation. The points above the line are inconsistent with our model. Observed values of these parameters are shown with error bars.
    }
    \label{fig:Bayes_avse}
\end{figure}

In terms of Bayesian statistics, a PTA produces a posterior on $\ayr$,
\begin{equation}
    p \left( \ayr | {\rm PTA} \right) \propto p \left( \ayr \right) ~ p \left( {\rm PTA} | \ayr \right), 
\end{equation}
where $p \left( \ayr \right)$ is the prior distribution, and $p \left( {\rm PTA} | \ayr \right)$ is the likelihood. We are interested in producing a posterior on parameters from our model. As the $M_{\bullet}$-$\mbulge$ relation provides the most influential parameters on $\ayr$, here we calculate the posteriors on $\alpha$, $\beta$, and $\epsilon$:
\begin{equation}
    p \left( \alpha, \beta, \epsilon | {\rm PTA} \right) \propto p \left( \alpha \right) ~ p \left( \beta \right) ~ p \left( \epsilon \right) ~ p \left( {\rm PTA} | \alpha, \beta, \epsilon \right).
\end{equation}
Our model gives us a way of translating $\alpha$, $\beta$, and $\epsilon$ into a value of $\ayr$, which means the two likelihoods are equivalent,
\begin{equation}
    p \left( {\rm PTA} | \alpha, \beta, \epsilon \right) = p \left( {\rm PTA} | \ayr (\alpha, \beta, \epsilon) \right).
\end{equation}
We do need to include other observational parameters into our model, specifically the GSMF and the galaxy merger rate. We will represent all of these parameters with $\theta$, which we can marginalize over, giving
\begin{equation}
    p \left( \alpha, \beta, \epsilon | {\rm PTA} \right) \propto \int d\theta ~p(\theta) ~p(\ayr ( \alpha, \beta, \epsilon, \theta ) | {\rm PTA}).
\end{equation}
Fig.~\ref{fig:TriPlot_abe} shows the translation of the recent upper limit from \citet{ppta+15} into the parameter space that characterizes the $M_{\bullet}$-$\mbulge$ relation. We can set a $95\%$ upper limit in the $2-D$ parameter space, $\alpha$-$\epsilon$, by integrating the distribution, shown in Fig.~\ref{fig:Bayes_avse}. 
Similar work has been done in \citet{middleton+16}, where an attempt is made to reconstruct a parametrized form of the black hole merger rate density from a posterior on $\ayr$ from PTAs. This kind of mapping from PTA data to astrophysical parameters is the next step forward in analyzing PTA data and is already being used in recent PTA limit papers \citep{nanograv+15}.

\subsection{Reconciling PTA limits with \mmb\ measurements}\label{sec:stall}

\begin{figure}
    \centering
    \includegraphics[width=\columnwidth]{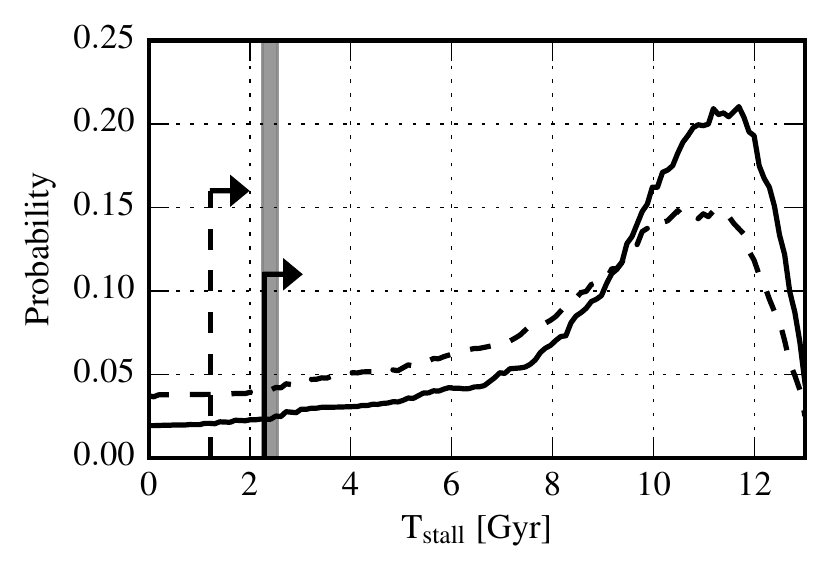}
    \caption{Translation of the marginalized posterior distribution of $\ayr$ using the measured parameters of the $M_{\bullet}$-$\mbulge$ relation from \citet{kormendyho13} and \citet{mcconnellma13} into a probability distribution of $T_{\rm stall}$, the time between the galaxy merger and the binary SMBH entering the PTA band in Gyrs. We set a lower limit of 1.2 and 2.3\,Gyr for the $\alpha, \epsilon$ values of \citet{mcconnellma13} and \citet{kormendyho13}, respectively. The shaded gray region is upper limit on the amount of $T_{\rm stall}$ expected from the simulations in \citet{khan+13}.
    }
    \label{fig:Tstall_UpLim}
\end{figure}

Fig.~\ref{fig:Bayes_avse} demonstrates that the \citet{kormendyho13} and \citet{mcconnellma13} measurements are inconsistent with the upper limit on the GWB by \citet{ppta+15}. We can thus attempt to reconcile this discrepancy by considering whether a non-zero $\tstall$ can make these two results consistent. If we assume the measured values are correct, then we can do a direct translation of the PPTA upper limit into a probability distribution on $T_{\rm stall}$. This is seen in Fig.~\ref{fig:Tstall_UpLim}, and we set a lower limit of 
$\tstall>2.3$\,Gyrs using \citet{kormendyho13} and a lower limit of
$\tstall>1.2$\,Gyrs using \citet{mcconnellma13}. 

We reiterate that in our formulation, $\tau$ essentially represents a dynamical friction timescale (the merger timescale), while $\tstall$ represents the additional time it takes the binary to enter the PTA waveband. Our limits on $\tstall$ could be capturing a longer merger timescale, such as those discussed in \citet{ravi+15}. However, since there is not a concrete definition for each of these parameters in the literature, we choose to make the straight-forward assumption that after a standard dynamical friction timescale estimate, the remaining time for binary evolution is captured by $\tstall$.

There are a limited number of observational and theoretical limits on stalling timescales in the literature. Originally, the apparently ``missing'' inspiral mechanisms were taken to imply that binary SMBHs might stall for up to a Hubble time \citep{begelman80}, although for a few systems with sufficient gas it was shown that binary SMBHs can inspiral efficiently \citep[\eg][]{mayer+07,cuadra+09}. More recently, numerical simulations have demonstrated that with mild triaxiality or axisymmetries in merging galaxies, binary SMBH coalescence timescales can be pushed to $\lesssim 2.4$\,Gyr, and contrary to previous inference, timescales have been found to be much less ($<0.5\,$Gyr) for the highest-mass galaxies \citep{preto+11,khan+11,khan+13}. Furthermore, \citet{radiocensus} placed an observational upper limit on the stalling timescale of the most massive binary SMBH systems ($\gtrsim10^8\,\msun$) of $<1.25$\,Gyr at 50\% confidence (i.\,e.\ a few Gyrs at 95\% confidence). The latter observationally-derived value is consistent with our lower limits using both the \citet{mcconnellma13} and the \citet{kormendyho13} relations. However, the theoretical result that inspiral due to host axisymmetries can be $\lesssim 2.4$\,Gyr, or beyond $<0.5$Gyr for the highest-mass galaxies we're probing here, is in contention with both of these limits. More recent work has shown that for `dry' mergers, where there is little gas present, the total merger time is $>1$ Gyr \citep{Sesana+15}, which is consistent with our limits.

There are two likely interpretations of this result. First, if we assume that the $\mbulge$ relations here hold, we must infer that either something else is reducing the expected GW background (for instance, a turn-over due to environmental coupling or a longer merger timescale than assumed here), or that triaxiality and axisymmetry are not prevalent enough to be the dominant force in driving pre-GW-dominant inspiral in massive galaxies.
Second, it is possible that the parameterizations of $\mbulge$ relations have been over-estimated.
As a more general consideration for this analysis, it is worth noting here that it has been suggested that recent black hole - host relations might err on using an upper-limit mass value for many black holes in their fits, for a range of valid values \citet[see e.\,g.\ the discussion in Chapter 3,]{merrittbook}. Thus, it is possible that the \citet{mcconnellma13} and \citet{kormendyho13} measurements have arrived at disproportionately high values, particularly for $\alpha$ and $\epsilon$. Recent work has proposed there is a bias in these measurements as well \citep{Shankar+16}, and we are currently working to assess what impact this effect would have on our results, we note that a moderate downward shift of both the \citet{mcconnellma13} and \citet{kormendyho13} relations in $\alpha$ and/or $\epsilon$ would render fits consistent with current PTA limits (e.\,g.\ Fig.\,\ref{fig:Bayes_avse}).

As previously noted, there could exist a deviation from a single-power-law GWB spectrum due to essentially the opposite effect from stalling: a super-efficient evolution through low orbital frequencies that also affects the PTA band. This may lead to a low-frequency turn-over in the strain spectrum and significant additional complexity to PTA analysis \citep[\eg][]{ravi+14,huerta+15,sampson+15,nanograv+15}. An investigation of that possibility is beyond the scope of this paper.

\section{Conclusions}

We inspected the amplitude variance of the nHz-$\mu$Hz waveband GWB, as formulated from state-of-the-art observations of massive galaxies, massive galaxy mergers, and SMBHs in the $z\lesssim3$ Universe. 
In the course of this analysis we provided simple reference plots which can be used to map PTA limits on the GWB to the parameters of cosmological evolution that most impact the power-law GWB prediction (Figs.\,\ref{fig:hvse}--\ref{fig:hvsTstall}).

We found that the vastly different observations of the intercept ($\alpha$) and scatter ($\epsilon$) of the $\mmb$ relation have the most impact on the range of GWB amplitude prediction.
We used the most constraining PTA upper limit on the GWB of $\ayr<1.0\times10^{-15}$ from \citet{ppta+15} to compare with our mapping to these parameters, and found that the $\mmb$ relations of \citet{mcconnellma13} and \citet{kormendyho13} are inconsistent with the PTA limit. Both of these measurements include the high-mass SMBHs expected to contribute the majority of signal to the PTA gravitational wave band \citep{kormendyho13,mcconnellma13}, making them the appropriate measurements for this comparison.

As discussed in Sec.\,\ref{sec:stall}, this inconsistency can be reconciled in a number of ways. First, we can include a moderate amount of ``stalling'' in the inspiral of the binary SMBH, in which the pair slows its evolution for at least 1.2 and 2.3\,Gyr for the $\alpha, \epsilon$ values of \citet{mcconnellma13} and \citet{kormendyho13}, respectively, which are in contention with theoretical work that has shown how axisymmetries may allow inspiral efficiencies of $<0.5$Gyr for the most massive pairs.

Observationally, the uncertainty in $\mmb$ and similar relations are largely due to small sample sizes. One important conclusion drawn from this analysis is that better constraints on the $\mmb$ relation
will greatly tighten GWB predictions, motivating further SMBH measurements to be made in the $\gtrsim10^9\,\msun$ range. 
Furthermore, it has been pointed out that the masses used for many SMBH - host galaxy relations may be overestimated, potentially solving the discrepancy we have found,  as recently noted in \citet{Shankar+16}. However, this only points more strongly towards a need for better characterization of these relations to allow an accurate prediction of the GWB, which we intend to pursue in more detail in future work.
The host-galaxy relation is most critical to be improved if we are to tighten the constraints PTAs can put on effects such as binary stalling, wandering SMBHs, and environmental interactions from the influence of gas or stellar dynamics during the GW-dominant regime. 
Finally, while the redshift evolution of this relation was not considered in this work, it is clear that strong evolution could further heighten its impact on GWB predictions (see, \eg, \citealt{ravi+15}).

We conclude by reiterating several caveats. We have considered only a power-law GWB, in which the binary inspiral is decoupled from its environment throughout the nHz--$\mu$Hz GW band. However, a quantification of super-efficient evolution due to strong environmental coupling poses an equally large, if not much larger, source of uncertainty in the low-frequency end of the PTA band. This must be better assessed via observational and theoretical work.


\section*{Acknowledgements}
The authors thank the Parkes Pulsar Timing Array for providing the posterior distribution on $\ayr$ used in Section\,\ref{sec:ptacon}.
Much acknowledgement goes to X.\ Siemens for extensive discussion of this work, A.\ Sesana for help in developing this project and C.\ Pankow for valuable technical assistance. 
The authors also acknowledge insightful comments from the anonymous referee, which helped improve this manuscript.
Sharelatex.com was used in the initial preparation of this manuscript.
JS is supported through the National Science Foundation (NSF) PIRE program award number 0968296 and NSF Physics Frontier Center award number 1430284. JS is partially supported by the Wisconsin Space Grant Consortium. SBS is a Jansky Fellow.
The online cosmology calculator of \citet{cosmocalc} was used as a reference in the course of this work.

\bibliographystyle{apj}
\bibliography{bibliography}

\end{document}